\documentclass[sigconf]{acmart}

\renewcommand\footnotetextcopyrightpermission[1]{}
\usepackage{booktabs} 

\usepackage{pifont}
\usepackage[english]{babel}
\usepackage[utf8x]{inputenc}
\usepackage{amsmath}
\usepackage{graphicx}
\usepackage{todonotes}
\usepackage{url}
\usepackage{subcaption}
\usepackage{diagbox}

\usepackage[export]{adjustbox}

\begin{document}

\copyrightyear{2017}
\acmYear{2017}
\setcopyright{rightsretained}
\acmConference{WebSci '17}{}{June 25-28, 2017, Troy, NY, USA}
\acmDOI{10.1145/3091478.3098879}
\acmISBN{978-1-4503-4896-6/17/06}
\acmPrice{}


\fancyhead{}
\settopmatter{printacmref=false, printfolios=false}

\title{Ongoing Events in Wikipedia: A Cross-lingual Case Study}

\author{Simon Gottschalk*, Elena Demidova*, Viola Bernacchi**, Richard Rogers***}
\affiliation{%
  \institution{*L3S Research Center, Hannover, Germany \\
  **DensityDesign Research Lab, Milano, Italy\\
  ***University of Amsterdam, Amsterdam, The Netherlands
  }
}
 \email{demidova@L3S.de, gottschalk@L3S.de,
  viola.bernacchi@live.it, R.A.Rogers@uva.nl}

\renewcommand{\shortauthors}{S. Gottschalk et al.}

\begin{abstract}
In order to effectively analyze information regarding ongoing events that impact local communities across language and country borders, researchers often need to perform multilingual data analysis. 
This analysis can be particularly challenging due to the rapidly evolving event-centric data and the language barrier. 
In this abstract we present preliminary results of a case study with the 
goal to better understand how researchers interact with multilingual event-centric information in the context of cross-cultural studies and which methods and features they use. 
\end{abstract}

\maketitle

\section{Introduction}
\label{sec:introduction}

In case of events concerning several communities, event-centric information available in different languages can reflect community-specific bias \cite{Rogers:2013}.  
As events unfold, it therefore becomes particularly important to analyze language-specific differences across the local event representations to better understand the community-specific perception of the events, the propagation of event-centric information across languages and communities as well as its impact.

Wikipedia with its large multicultural user community and editions in over 290 languages is a rich source of multilingual information with respect to ongoing and past events.
%
%
Information about ongoing events of global importance can evolve quickly in different language editions as the event unfolds \cite{Fetahu:2015}, making Wikipedia an important and up-to-date source for cross-cultural studies.

The potentially large amount of multilingual information to be analyzed, event dynamics and the language barrier
limit the analysis methods, in particular with respect to the features accessible for the analysis, making cross-cultural studies of ongoing events particularly challenging.  
Although existing tools facilitate automatic cross-lingual article comparisons in Wikipedia with respect to selected features, including for example a cross-lingual text alignment for articles pairs by MultiWiki \cite{Gottschalk2017}, a comparison of the topics covered by the articles by Omnipedia \cite{Bao:2012} and a computation of the article similarity in Manypedia \cite{Massa:2012}, there is still a significant room for further developments in this area. 

In this abstract we present preliminary results of a case study concerning an analysis of cross-lingual event-centric information 
in Wikipedia exemplified through the Brexit referendum in June 2016.
In this study, we observed an interdisciplinary, multicultural group of information professionals (communication design and sociology researchers) who performed the 
task of cross-cultural analysis of the Brexit representation in the multilingual Wikipedia in June 2016. In the study, six participants worked for 14 hours together as a team, resulting in a total of 84 person-hours spent. 
%

\section{Objectives \& Methodology}
\label{sec:challenges}
The key study objective was to
\textit{observe which methods and features can be efficiently used to perform cross-lingual analytics on multilingual Wikipedia}. 
As a first step, we introduced the participants to the topic of Brexit in Wikipedia and presented them with several example tools that can assist Wikipedia analytics in general. 
We asked the participants to: 1) define their own research questions and analysis methods related to Brexit and its international perception from the social and cultural perspectives; and 
2) conduct an analysis of the Brexit-related articles across
different Wikipedia language editions with respect to these questions.
We focus the discussion on two Brexit-related
articles selected by the participants: 
the “United Kingdom European Union membership referendum, 2016” (\textit{referendum}) article and the “United Kingdom withdrawal from the European Union” (\textit{withdrawal}) article.

\begin{figure*}
	\centering
	\includegraphics[width=0.9
    \textwidth]
    {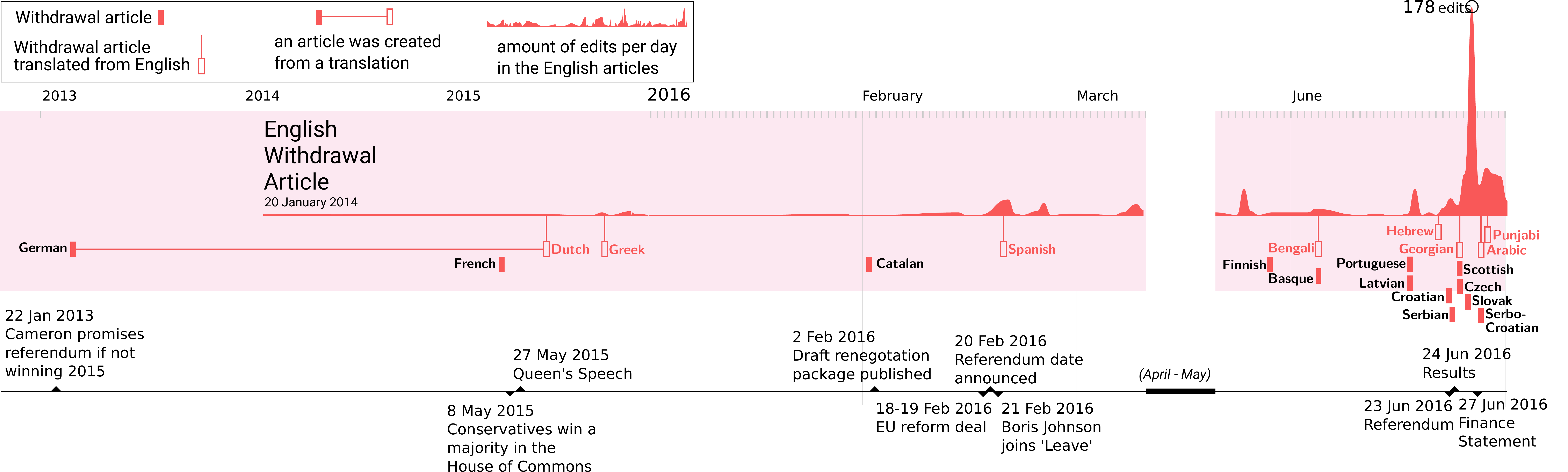}
	\caption{Timeline of the withdrawal article showing the edit frequency of the English withdrawal article over time and the creation of the article editions in other languages. 
    }
    \label{fig:timeline}
\end{figure*}

\section{Observations}
\label{sec:feature_selection}

\subsection{Content Analysis} 
Due to the language barrier, the participants focused on the content features involving less text, e.g. the tables of contents and images.

\textbf{Textual features}: Most Wikipedia articles contain a table of contents (TOC) that is arranged individually in each language edition.
As the effort to read through whole Wikipedia articles -- especially when the reader is not proficient in the particular language -- was estimated too high, the participants decided to focus on the TOCs as an approximation of textual analysis. 
%
%
The participants arranged the TOC entries from the English, German, French and Italian article revisions by their frequency across languages. This TOC comparison has shown that the articles differed in many aspects: While the German article focused on the referendum results in the different regions, the English Wikipedia covered much more aspects, especially on economical and societal implications of the withdrawal.

\textbf{Multimedia features}: Wikipedia articles can contain images that can be shared across different language editions. It became evident that images containing the map of the UK and the referendum ballot paper were shared most frequently across languages. Smaller deviations were observed for several languages, e.g. the German article contained many photos of the politicians.

\vspace{-1em}

\subsection{Temporal Analysis} The description of ongoing events may vary substantially over time. 

\textbf{Textual features:} The TOC of a Wikipedia article is subject of change over time. To observe this behavior, the participants extracted the TOC three times per day, in the time from the 22nd to the 24th of June 2016 (with the 23rd of June being the referendum date). 
The TOCs of two of these points in time made visible that the French version did not have the referendum results on the 23rd of June as opposed to other language versions under consideration. Within the following day, the English TOC stayed nearly unchanged in contrast to the other languages where especially the German article showed a major increase of new sections. 
The analysis on the withdrawal article revealed that the German version contained detailed historical and legislative entries while the French version focused on the referendum's impact and the public opinion.

\textbf{Edit-based features:} A valuable temporal aspect with respect to the cross-lingual information propagation comes from the editing process by the Wikipedia editors and the origin of the first revision per language. To explore this feature, the lists of all language editions for the referendum ($48$ language editions) and the withdrawal articles ($59$ language editions) have been analyzed using the Wikipedia Edits Scraper and IP Localizer tool. 
This way, the participants gained the edit histories of $107$ article language editions. 
They manually created two annotations for the initial revision of each article: “Creation date”, i.e. the date of the first edit, and “Origin”, i.e. the reference to other language editions that were utilized to create the respective edition. 

A visualization presented in Figure \ref{fig:timeline} has been created by the participants and illustrates the English articles’ development including the number of edits and the creation of the other language editions. Article editions directly translated from other languages are marked and important events related to Brexit are added to the timeline. 
Edit peaks correlate with the sub-events in particular during the referendum held on the 23rd of June 2016.

\subsection{Network \& Controversy Analysis} 
The interlinking between Wikipedia articles across languages can provide useful insights into the coherence of the language editions. 

\textbf{Category-based features:}
In Wikipedia, editors can assign categories like “Politics” or “EU-skepticism in the UK” to the articles. To gain more insights into the perception of Brexit within the network of Wikipedia articles, the participants analyzed this categorization 
in all available language editions.
The majority of the categories was related to basic concepts like “Referendum”. 
The Scottish and the English Wikipedia were rather disconnected from the network. In the withdrawal articles, “Eurocentrism” was identified as a main category connecting languages.

\textbf{Links-based features:} Each Wikipedia article can contain links to external sources that the participants extracted and compared using the MultiWiki tool. For most of the language editions, the overlap of the linked web pages was rather low 
, and just in some cases it reached higher values: the English and German withdrawal articles shared $17.32\%$ of links, mainly related to news pages like the BBC or The Guardian.

%
\textbf{Edit-based features:} To observe controversies, the participants reviewed the English discussion pages on the articles using their TOCs. A major finding facilitated by this feature was a discussion among the Wikipedia editors on the question to which article the search term “Brexit” should link to. Having initially redirected to the referendum article, the term lead to the withdrawal article a few days after the actual referendum. 

\section{Conclusion}
\label{sec:conclusion}

As our case study illustrates, certain types of cross-lingual analysis on Wikipedia 
mostly relying on non-textual features (e.g. links, images and category structures) 
can be performed efficiently, manually or by using existing tools, and can already provide interesting insights. In contrast, analysis of textual content and discussion pages appears to be limited through the language barrier. 

\footnotesize{
\section{Acknowledgment}
Partially funded by ALEXANDRIA
(ERC 339233) and Data4UrbanMobility (German Federal Ministry of Education and Research (BMBF) 02K15A040).

\bibliographystyle{ACM-Reference-Format}
\bibliography{references} 
}

\end{document}